\documentclass[twocolumn,prl,superscriptaddress]{revtex4-2}

\usepackage[normalem]{ulem}
\usepackage{amsmath}
\usepackage{amssymb}
\usepackage{graphicx}
\usepackage{dcolumn}
\usepackage{bm}
\usepackage{graphicx}
\usepackage{xcolor}
\usepackage{placeins}
\newcommand{\qq}{\begin{eqnarray}}
\newcommand{\qqq}{\end{eqnarray}}

\newcommand{\bfr}{\mathbf{r}}

\newcommand{\bfs}{\boldsymbol{\sigma}}

\newcommand{\bfv}{{\bf v}}

\newcommand{\bfQ}{{\bf Q}}
\newcommand{\bfA}{{\bf A}}
\newcommand{\bfO}{{\bf \Omega}}
\newcommand{\Fcal}{\mathcal{F}}

\newcommand{\pg}[1]{\textcolor{black}{#1}}

\newcommand{\mcm}[1]{\textcolor{black}{#1}}

\begin{document}

\preprint{APS/123-QED}

\title{
\mcm{Active fluids form system-spanning filamentary networks}}
\author{Paarth Gulati}
\affiliation{Department of Physics, University of California Santa Barbara, Santa Barbara, CA 93106, USA}
\author{Fernando Caballero}
\affiliation{Department of Physics, Brandeis University, Waltham, Massachusetts 02453, USA}
\author{M. Cristina Marchetti}
\affiliation{Department of Physics, University of California Santa Barbara, Santa Barbara, CA 93106, USA}
\affiliation{Interdisciplinary Program in Quantitative Biosciences, University of California Santa Barbara, Santa Barbara, CA 93106, USA}


\begin{abstract}
Recent experimental realizations of liquid-liquid phase separation of active liquid crystals have offered an insight into the interaction between phase separation, ubiquitous \mcm{in soft matter and biology}, and chaotic active flows. In this Letter, we \mcm{use continuum theory to examine phase separation of an active liquid crystal and a passive fluid and report two new results. First, we provide an analytical derivation of the activity-induced suppression of the phase boundary of the coexistence region - a result first reported in simulations and experiments.}
We show that the shift in the critical point is a result of the balance between self-stirring active flows and phase-separating diffusive fluxes.
Second, we show that this same balance is responsible for \mcm{dramatically changing the morphology of the phase separated state, resulting in}
the emergence of a new mixed active phase
consisting of a dynamical filamentous active network that invades  the entire system area, trapping droplets of passive material.  This structure exists even for very low volume fractions of active material. \mcm{Our work provides an important step towards the goal of understanding how to use activity as a new handle for sculpting interfaces.}
\end{abstract}
\maketitle

\mcm{The phases separation of two immiscible fluids is a ubiquitous phenomenon in materials science and has recently taken center-stage in biology with the discovery of membrane-less organelles~\cite{gomes2019molecular}. Experiments and simulations have begun to explore the role of activity on liquid-liquid phase separation (LLPS), demonstrating that sustained energy injection at the microscale - the hallmark of active matter - profoundly modifies the kinetics of fluid mixing and demixing.
Activity can give rise to new types of phase separation not possible in equilibrium~\cite{tailleur2008statistical,fily2012athermal,Redner2013,stenhammar2015activity, bhattacharyya2023phase}, and it can arrest~\cite{tiribocchi2015active, wittkowski2014scalar} and suppress~\cite{caballero2022activity,tayar2023controlling} phase separation. It also gives rise to a wealth of new interfacial phenomena, including giant interfacial fluctuations, traveling surface waves in the absence of inertial restoring forces, and unusual wetting behavior~\cite{adkins2022dynamics,gulati2024traveling,zhao2024asymmetric}. }

\mcm{The role of activity on LLPS has been explored experimentally in active liquid crystals of microtubule-kinesin bundles that exert extensile forces on their environment. In bulk these fluids are generically unstable ~\cite{aditi2002hydrodynamic} and spontaneously generate self-sustained chaotic flows~\cite{sanchez2012spontaneous,alert2022active}. When immersed} in a mixture of two immiscible polymers,
the active microtubule bundles preferentially partition to one of the two species,
\mcm{thus yielding a phase separating active/passive mixture}~\cite{adkins2022dynamics, tayar2023controlling,zhao2024asymmetric}.

\mcm{A fruitful} approach for a theoretical understanding of these \mcm{phenomena} are continuum theories \mcm{that modify well-established $\phi^4$ or Cahn-Hilliard models of LLPS to couple to liquid crystalline degrees of freedom and incorporate active processes that drive the system out of equilibrium.} It was recently shown~\cite{caballero2022activity,tayar2023controlling} that activity can alter the equilibrium phase diagram, effectively renormalizing the temperature-like parameter that \mcm{controls the preference for} mixing or phase separation.
The properties of the coexistence \mcm{region} remain, however, largely unexplored. Likewise, \mcm{the study of the role of activity on the  shape and geometry of interfaces has so far been mainly limited to }single droplets~\cite{blow2014biphasic, tiribocchi2015active,xu2023geometrical,kempf2019active}. \mcm{Understanding how active energy injection modifies the morphology of the phase separated state is an important open question, with direct relevance to the experimental ambition of using activity as a new tool for sculpting fluid-fluid interfaces.}

In this Letter, we \mcm{employ} a continuum theory \mcm{to examine phase separation in a} mixture of an active liquid crystal and an isotropic passive fluid. \mcm{We present an analytical derivation of the dependence on activity} \mcm{of the phase boundary delimiting the coexistence region, which was previously reported in simulations~\cite{caballero2022activity} and experiments~\cite{tayar2023controlling}.} \mcm{We additionally demonstrate that} extensile active stresses \mcm{dramatically} change the morphology of \mcm{the phase separated state. Unlike in equilibrium, where the minority phase forms }spherical droplets, \mcm{the active fluid} forms a single continuous filamentary network that percolates the entire system, trapping droplets of the passive phase within the open spaces of the network. \mcm{Remarkably,} this happens even for very low volume fractions of the active phase. \pg{While in this Letter we focus on two-dimensions, we note that the morphology of the phase separated fluid depends on the dimensionality of space. In three dimensions, preliminary results show that for all compositions active and passive fluids phase separate into a bicontinuous structure at steady state~\cite{SI}.}

\emph{\mcm{The model.}} We consider \mcm{an established} continuum model for a mixture of an active liquid crystal and a passive fluid~\cite{blow2014biphasic, zhao2024asymmetric}. The relative concentration of active and passive material is described by a phase field $\phi(\bfr, t)$ that obeys Cahn-Hilliard dynamics and is advected by the flow velocity $\bfv(\bfr,t)$, 
\begin{equation}\label{eq:continuityphi}
\partial_t\phi + \bfv\cdot\nabla\phi = M\nabla^2\mu\;,
\end{equation}
where \mcm{$M$ is a mobility} and $\mu=\delta \Fcal/\delta\phi$ is the chemical potential, with 
$\mu = a\phi+|a|\phi^3 - k\nabla^2\phi$. We 
 consider $a<0,$ which
 promotes phase separation between bulk phases $\phi=\pm 1$, with interfaces of width $\xi=\sqrt{k/|a|}$, and an interfacial tension $\sigma=\sqrt{8k|a|/9}$~\cite{cates2018theories}. We will use the convention that $\phi=1$ and $\phi=-1$ describe active and passive phases respectively.

A traceless tensor field $\bfQ$ describes the liquid-crystalline degrees of freedom of the active nematic liquid crystal, with dynamics given by~\cite{marchetti2013hydrodynamics, ramaswamy2010mechanics}
\begin{equation}\label{eq:Qdynamics}
    \partial_t \bfQ  +\bfv\cdot\nabla\bfQ = \bfQ  \cdot \boldsymbol{\Omega}-\boldsymbol{\Omega}\cdot \bfQ  +\lambda \mathbf{A} + \frac{1}{\gamma}\mathbf{H}\;,
\end{equation}
where $\bfA$ and $\bfO$ are the strain rate and vorticity tensors, with components $A_{ij} = (\partial_i v_j + \partial_j v_i)/2, \Omega_{ij}=(\partial_i v_j - \partial_j v_i)/2 $. The molecular field $\mathbf{H}$ describing equilibrium relaxation, with $\gamma$ a rotational friction, \mcm{is obtained} from a Landau-de Gennes free energy, \mcm{with the result}
\begin{align}\label{eq:mol_field}
    \mathbf{H} &= -r\Tilde{\phi}\bfQ  -u\rm{Tr }{\bfQ^2}\bfQ  + K \nabla^2 \bfQ \;.
\end{align}
where $\Tilde{\phi} = (1+\phi)/2$ is the volume fraction of the active fluid.

Finally, the velocity field is \mcm{governed by force balance} in the Stokes' regime,
\begin{equation}\label{eq:momentum_eq}
    0 = \eta \nabla^2 \mathbf{v} + \nabla\cdot \boldsymbol{\sigma} -\nabla P\,
\end{equation}
where the pressure $P$ is determined by incompressibility $\nabla\cdot\bfv = 0$. The  stress tensor $\bfs$ \mcm{is the sum of} the following contributions
\begin{equation}\label{eq:stress}
    \boldsymbol{\sigma} = \alpha \Tilde{\phi}\bfQ  + \boldsymbol{\sigma}^{c} + \boldsymbol{\sigma}^{e}.
\end{equation}
The first contribution represents the active stress, where $\alpha<0$ ($\alpha>0$) describes an extensile (contractile) liquid crystal. In this work, we restrict ourselves to the extensile case. The second and third contributions are the equilibrium capillary \mcm{stress}, $\sigma^c_{ij} = -k \left[(\partial_i \phi) (\partial_j \phi) - (\nabla\phi)^2\delta_{ij}/2\right]$ and liquid crystalline elastic stress, $\sigma^{e}_{ij} = -\lambda H_{ij}+ Q_{ik}H_{kj} - H_{ik}Q_{kj}$ \cite{cates2018theories,cates2019active,  marchetti2013hydrodynamics}.

To explore the behavior of the system, we numerically integrate Eqs.~(\ref{eq:continuityphi}-\ref{eq:momentum_eq}) using self-developed pseudospectral solvers~\cite{caballero2024cupss} on a regular, periodic, two-dimensional grid of size $L\times L$. The parameter values are reported in time units of $\gamma/u$, lengths units of $\sqrt{M\gamma}$, and energy density in units of $u$. Additionally, we fix the equilibrium interface width $\xi=\sqrt{3/2}$, viscosity $\eta=1.0,$ and the flow alignment parameter $\lambda=1.0$.

To explore the phase diagram of the system, we tune the equilibrium interfacial tension $\sigma$, activity $\alpha$ and mean composition of the mixture $\phi_0=\int_V d\bfr \phi(\bfr,t)$. We consider the liquid crystal to be in the isotropic phase at equilibrium, i.e., $r>0$, in line with the recent  \emph{in-vitro} assays of active MTs \mcm{that inspire} our work 
\cite{henkin2014tunable, adkins2022dynamics, zhao2024asymmetric}. \mcm{In this case} nematic \mcm{order}
is driven by active flows, which are \mcm{spontaneously} generated above the critical activity $\alpha_c = 2\eta r/\lambda \gamma$ where the \mcm{homogeneous} isotropic \mcm{state is}
 linearly unstable. \cite{voituriez2005spontaneous, giomi2011excitable, giomi2012banding, giomi2014spontaneous}.  
For the rest of this work, we fix $r=0.1$ i.e. $\alpha_c =0.2$. 
\mcm{Results for an ordered nematic base state with $r<0$ are presented in the SI.}
\\
\begin{figure}[t]
    \centering
    \includegraphics[width=0.95\linewidth]{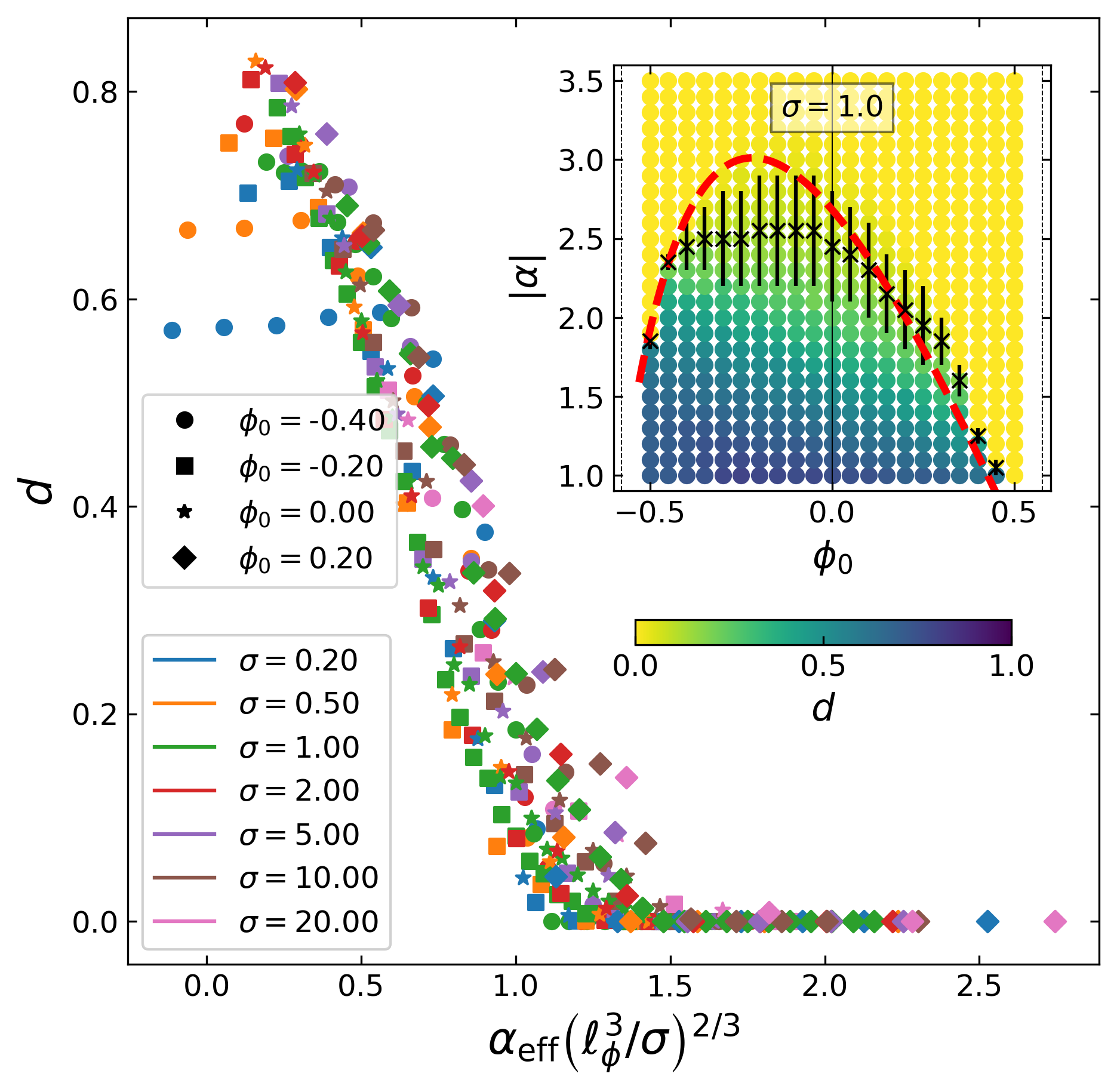}
    \caption{\textit{Active Mixing in the isotropic liquid crystal} ($r>0$)  The mean, steady-state demixing order parameter $d$ \mcm{is shown} as \mcm{a} function of $\alpha_{\rm{eff}} (\ell_\phi^3/\sigma)^{2/3}$ for varying surface tensions $\sigma$ (colors) and mean compositions $\phi_0$ (shapes). Here $\alpha_\text{eff} = |\alpha|\Tilde{\phi}_0-\alpha_c$ is the \mcm{excess} extensile activity of a uniform mixture of composition $\phi_0$ \mcm{relative to} the 
    \mcm{instability threshold} $\alpha_c$, with $\Tilde{\phi}_0 = (1+\phi_0)/2$  the mean volume fraction of the active fluid, and $\ell_\phi$ is the length scale of the most unstable mode driving the spinodal instability.
    Inset: \mcm{Phase diagram in the $(|\alpha|,\phi_0)$ plane for surface tension $\sigma=1.0$. The color is the} steady state mean $d$.
    The black \mcm{error bars} (\mcm{of range $d=(0.1-0.4)$}) denote the mixing transition in activity as a function of $\phi_0$. The dashed red line \mcm{is the fit obtained with}
    Eq.~\eqref{eq:alpha_mix_isotropic}, \mcm{with} $s_1 =  2.84\pm0.46, s_2 = 0.17 \pm 0.10.$ The phase diagram is explored within the spinodal region i.e. $-1/\sqrt{3} <\phi_0<1/\sqrt{3}.$
    Simulation Parameters: $L=256, K=6.0$.}
    \label{fig:mixing_collapse_isotropic}
\end{figure}

\emph{\mcm{Activity shifts the coexistence line.}} We first discuss the \mcm{shift of the critical point and coexistence line driven} by active mixing, originally described in \cite{caballero2022activity}, where it was observed that sufficiently strong active stresses suppress the equilibrium phase separation.  To quantify the amount of mixing in the system we define a demixing order parameter \mcm{as the spatial variance of the concentration field,}
\begin{equation}
\label{eq:d}
    d = \dfrac{\langle \phi^2\rangle - \phi_0^2}{1-\phi_0^2}\;,
\end{equation}
where $\langle \,\cdots \rangle$ \mcm{denotes a} spatial \mcm{average}.  For a phase separated mixture (with vanishingly thin interfaces) $d\rightarrow 1$, while for a homogeneous system $d \rightarrow 0$. We report the \pg{time averaged} values of the demixing parameter $d,$ averaged over the dynamical steady state. We show the structure of the phase diagram in Fig.~\ref{fig:mixing_collapse_isotropic} (inset). \mcm{Large activities suppress} demixing  and the critical point is shifted to values of the volume fraction $\phi_0$ lower than the equilibrium value $\phi_0=0$. 

To understand the mechanism of this \mcm{phenomenon}
and the origin of the 
{shift of the critical point,} 
we consider the growth rate of fluctuations of  length $\ell$ about a homogeneous state. 
\mcm{Hydrodynamic flows and diffusion tend to demix, while}  active flows stir the system. The net growth rate \mcm{of $\ell$} can be written as the sum of these rates as~\cite{caballero2022activity}
\begin{equation}\label{eq:lsold}
    \frac{\dot{\ell}}{\ell} = c_1 \frac{M \sigma}{\ell^3} +c_2\frac{\sigma}{\eta \ell} - c_3 \frac{S \alpha_\text{eff}}{\eta},
\end{equation}
where $c_i$ are unknown dimensionless constants and $S=\sqrt{\text{Tr}\{\bfQ^2\}}$ is the mean scalar order parameter. 
\mcm{The last term in Eq.~\eqref{eq:lsold} is the rate of active mixing, which is controlled by} the effective activity of a homogeneous mixture of composition $\phi_0$, \mcm{given by}  $\alpha_\text{eff} = |\alpha|\Tilde{\phi}_0-\alpha_c$, with $\Tilde{\phi}_0 = (1+\phi_0)/2$ the mean volume fraction of the active fluid.

In the absence of activity (within the spinodal region), the initial growth rate of the phase separated domains is controlled by the fastest-growing, linearly unstable mode characterized by the spinodal decomposition length, $\ell_\phi={\xi}/{\sqrt{1-3\phi_0^2}}.$ The spinodal length depends on the mean composition of the system and diverges as $\phi_0 \rightarrow \pm \phi_{\text{sp}}$, where $\phi_{\text{sp}} = 1/\sqrt{3}$.
\mcm{We can then identify the onset of activity-driven mixing with the vanishing of the} growth rate \mcm{of} the spinodal length scale, \mcm{corresponding to} $\dot{\ell}/\ell \rightarrow 0$ for $\ell = \ell_\phi$. \mcm{We can then estimate the characteristic activity $\alpha_{\rm{mix}}$ required for mixing from Eq.~\eqref{eq:lsold} as}
\begin{equation} 
\alpha_{\rm{mix}}\Tilde{\phi}_0
-\alpha_c 
= \left[\sigma \left(\frac{s_1 M }{\ell_\phi^3} + \frac{s_2}{\eta \ell_\phi}\right) \eta \sqrt{u} \right]^{2/3}\;, \label{eq:alpha_mix_isotropic}
\end{equation}
where 
$s_1 = c_1/c_3$, $s_2=c_2/c_3$, and we have approximated $S \sim \sqrt{\alpha_{\text{eff}}/u}$ as activity \mcm{drives nematic order by} effectively renormalizing  $r$ \cite{giomi2012banding, santhosh2020activity}.
%
\begin{figure}[t]
\includegraphics[width=\linewidth]{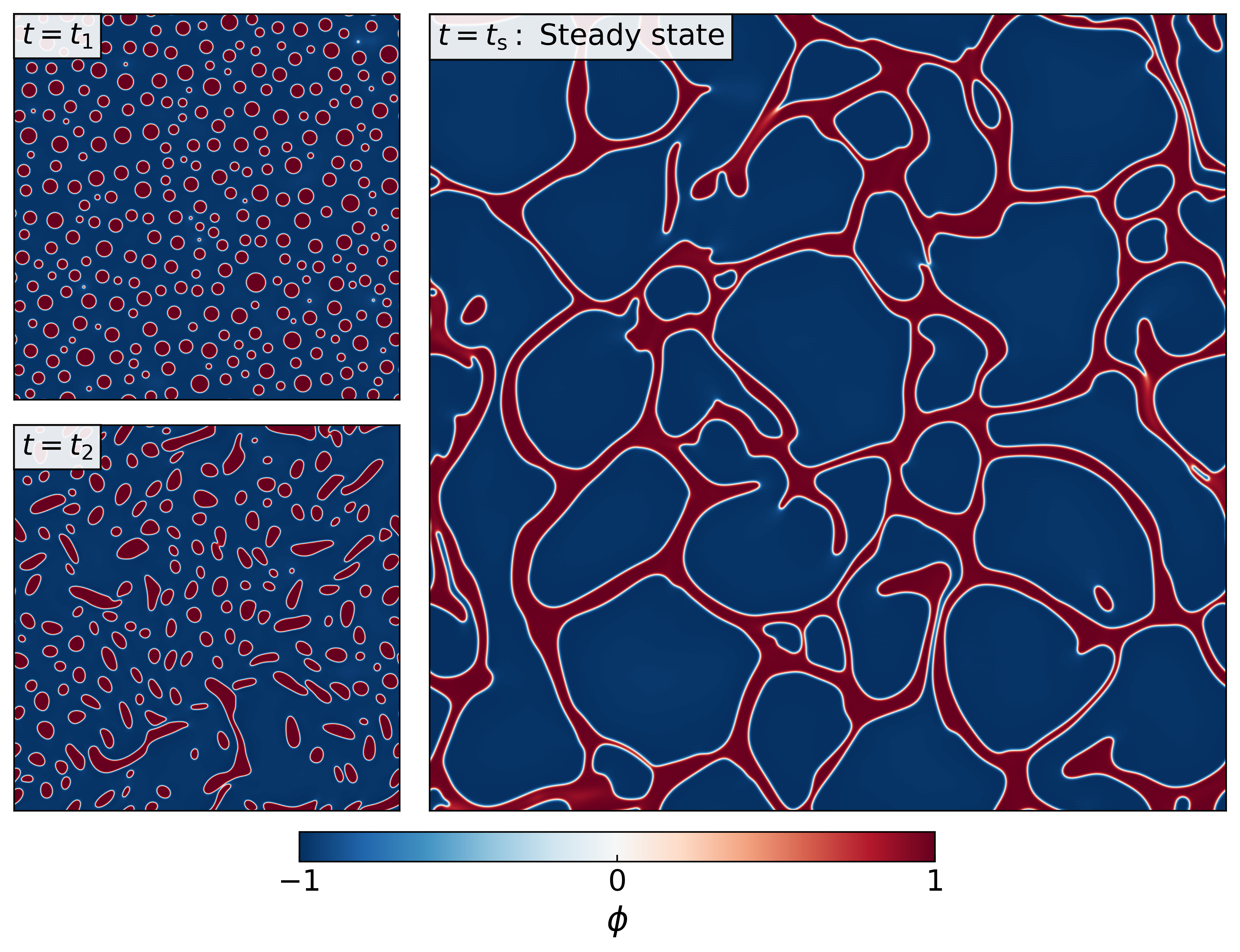}
\caption{Snapshots showing the generic pathway \mcm{to} the formation of a dynamical active network from an initially perturbed uniform state. Within the spinodal region
, concentration fluctuations lead to the formation of active ($\phi=1$) droplets ($t=t_1$). The droplets grow through Ostwald ripening. When they are sufficiently large, they acquire nematic order in the interior, which in turn leads to extensile flows that elongate the droplets ($t=t_2$). Eventually the elongated droplets merge with each other, forming a dynamical steady state consisting of a filamentous active network ($t=t_s$). Simulation parameters: $\phi_0=-0.55, \sigma =1.0, \alpha=-1.0, K=24.0$, and $L=1024$.}
\label{fig:percolation_timeSeries}
\end{figure}
 At short time scales (and away from $\phi_{\rm{sp}}$), diffusive processes dominate over hydrodynamic flows, 
and Eq.~\ref{eq:alpha_mix_isotropic} reduces to $\alpha_{\rm{mix}}\Tilde{\phi}_0-\alpha_c\sim (\sigma/\ell_\phi^3)^{2/3}$,
\mcm{with $\alpha_{\rm{mix}} \sim \sigma^{2/3}$
at large \mcm{activity} $\alpha_{\rm{mix}} \gg 
\alpha_c/\Tilde{\phi}_0$}. 
As shown in Fig.~\ref{fig:mixing_collapse_isotropic}, this is in \mcm{excellent} agreement with the numerical phase boundary for a wide range of compositions $\phi_0$ and surface tensions $\sigma$. 

\pg{The shift of the phase boundary in the activity/composition plane is given by the maximum of $\alpha_{\rm{mix}}$ given in Eq.~\eqref{eq:alpha_mix_isotropic} with respect to $\phi_0$. 
For large surface tension, $\alpha_{\rm{mix}}\Tilde{\phi_0} \gg \alpha_c,$ and $s_1 \gg s_2,$, and the location of the maximum is $\phi_{0, c} \rightarrow -1 + \sqrt{2/3} \approx -0.184$. This is in good agreement with our numerical results at large $\sigma$ \cite{SI}.
}

\mcm{Finally, the above discussion refers to the situation where the liquid crystal is isotropic when passive ($r>0$) and} nematic order \mcm{only emerges due to active flows (hence $S\sim \pg{\sqrt{\alpha_{\rm{eff}}/u}}$).
Similar arguments can be developed for the nematic phase ($r<0$). In this case \pg{$S \approx \sqrt{-r/u}\sim \rm{const.}$} and}
$\alpha_{\rm{mix}}\sim\sigma$, with $\phi_{0,c} \approx -0.121$~\cite{SI}.
\\

\emph{\mcm{Morphology of the phase separated state.}}
Activity strongly changes the morphology of the phase separated fluids in the demixed region. The steady state consists of a dynamical network of long elongated active filaments. This morphology is driven by extensile active flows created by the liquid crystal which aligns with the interface through active anchoring \cite{blow2014biphasic, coelho2023active, zhao2024asymmetric}. This single continuous network of active fluid \mcm{spans} the entire system, \mcm{with an emulsion of passive droplets occupying the interstitial regions. This filamentary structure is obtained} even for very low volume fractions of active material. \mcm{In this limit the majority}  passive phase  forms large droplets \mcm{(see Fig.~\ref{fig:percolation_timeSeries})}. As we increase activity, and the system approaches the phase boundary, mixing increases, and the network fractures into a microphase-separated steady state of elongated droplets that constantly merge and split.

\begin{figure}[t]
    \centering    \includegraphics[width=0.90\linewidth]{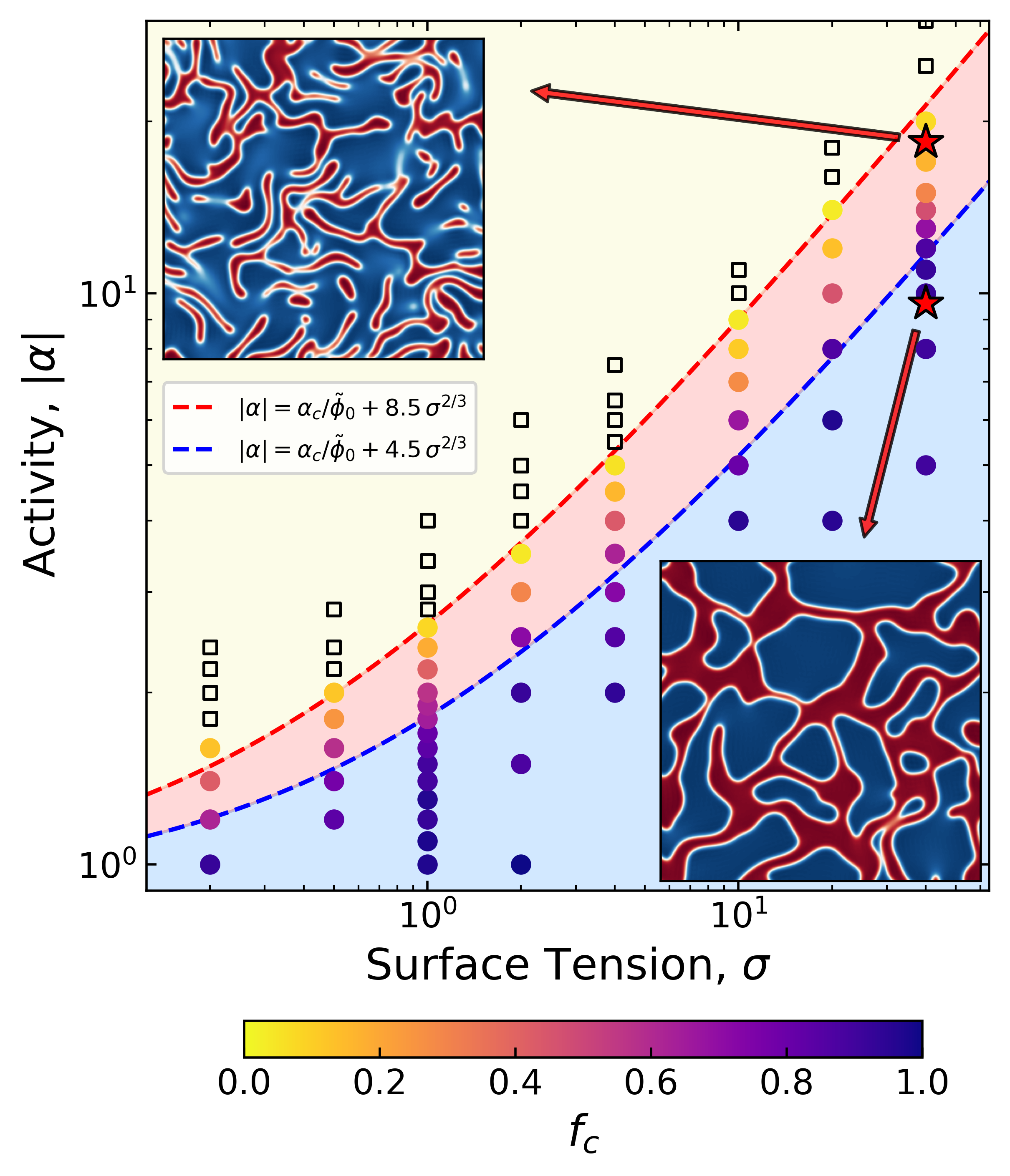}
    \caption{\textit{Morphology of demixed state.} Phase diagram as a function of activity $|\alpha|$ and surface tension $\sigma$.
    The filled-in circles \mcm{correspond to
    phase separated 
    steady states, with their color the value of $f_c$ as from the color bar.}
    The unfilled squares \mcm{correspond to well mixed}
    steady states ($d<0.1$).
    The dashed lines 
    \mcm{are fits to} $|\alpha| - \alpha_c/\Tilde{\phi}_0 = c\,\sigma^{2/3}$, with a single fitting parameter $c$ \mcm{(value given in figure legend).}
    The blue dashed line separates the filamentary network state \mcm{(light blue)} from phase separated droplets \mcm{(pink) and}
    is determined by using $f_c=0.7$ at $\sigma=1.0$. The red dashed line 
    \mcm{delimits the onset of the homogeneous well-mixed state (yellow), with $c$ determined by requiring }
    $d=0.1$ at $\sigma=1.0.$ 
    Two snapshots show representative steady states.
    Simulation parameters: $\phi_0=-0.55, K=24.0$, and $L=256$.}
    \label{fig:sigmaalpha}
\end{figure}

To \mcm{quantify the properties of the active network, 
we compute the volume fraction $f_c$} of the largest connected active component, \mcm{defined as} 
    $f_c = A_0/(\Tilde{\phi}_0 L^2)$,
where active regions are \mcm{those where} $\phi>0$ and $A_0$ is the area of the largest connected active domain. \mcm{The quantify $f_c$ provides} an order parameter \mcm{for the existence of the filamentous active structure:}  a single connected active region corresponds to $f_c\rightarrow 1$, while the microphase-separated states close to the phase boundary are characterized by $f_c\ll1$. In Fig.~\ref{fig:sigmaalpha} we use  $f_c$ as a metric to construct a phase diagram as a function of activity and surface tension.  
\mcm{In the light blue region bounded by the blue dashed line the active fluid forms a system-spanning network and $f_c$ is close to $1$. In the yellow region above the dashed red line the system is homogeneous with $d<0.1$ and $f_c$ close to zero. In the intervening pink region the network fractures in an increasing number of small droplets with increasing activity, and $f_c$ decreases continuously. The blue and red dashed lines are one-parameter fits to Eq.~\eqref{eq:alpha_mix_isotropic}, as described in the caption.}

\begin{figure}[b]
    \centering
    \includegraphics[width=0.99\linewidth]{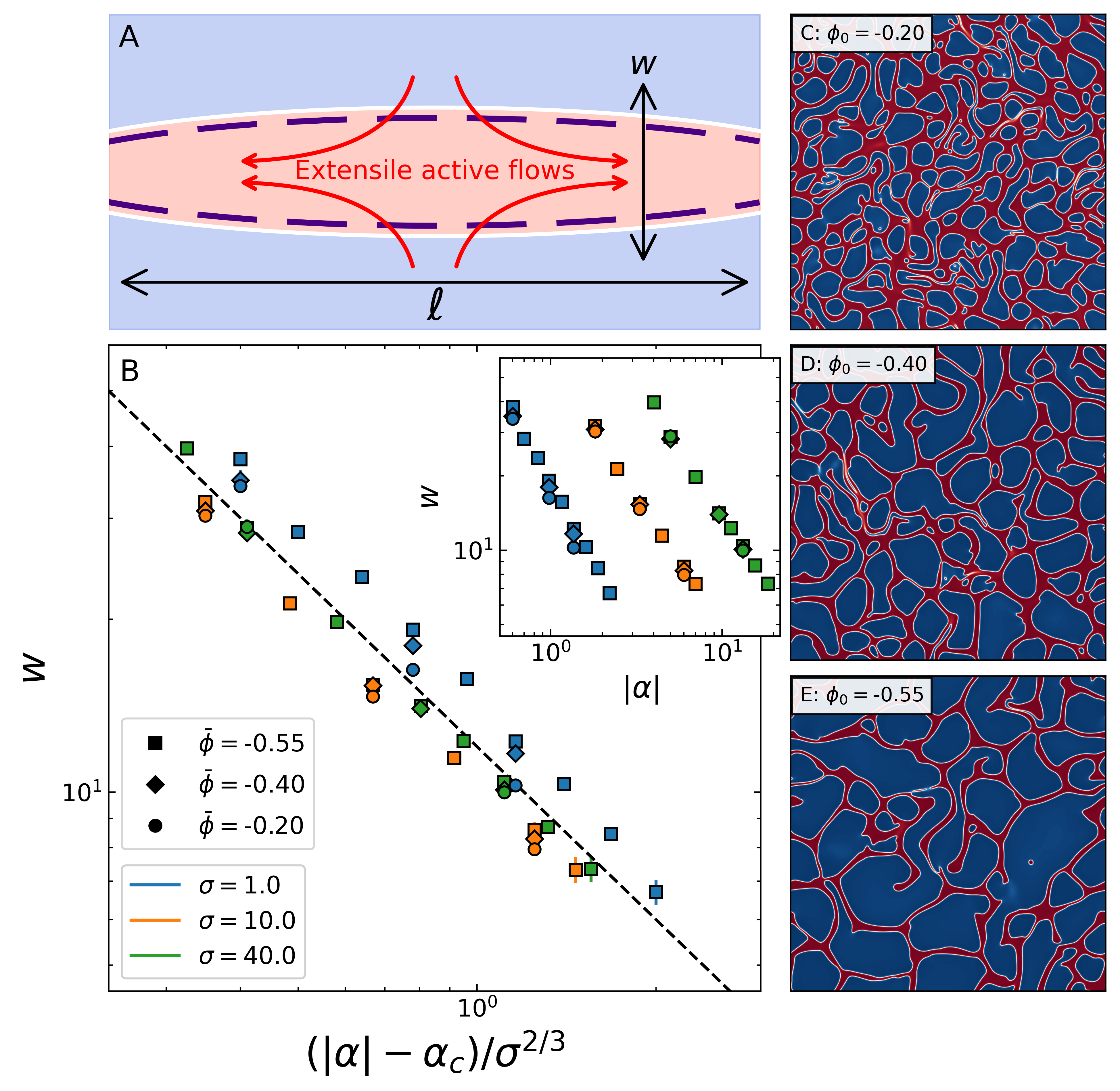}
    \caption{\textit{\mcm{Width of filamentary active network.}} 
    \mcm{A:} Sketch \mcm{describing how}  extensile active flows \mcm{elongate} active droplets due to effective anchoring of the director parallel to the interface.
    B: Mean steady state width of the \mcm{filamentary active network} computed from numerical simulations (using the skeletonization \mcm{described in the SI}), as a function of $(|\alpha|-\alpha_c)/\sigma^{2/3}.$ The dashed black line has  slope -1. Inset: \mcm{Network width as a function of unscaled} activity \mcm{for three mean concentrations}. 
   C-E: Representative snapshots of the steady state active network for three mixture compositions $\phi_0$, $\alpha = -1.0$ and $\sigma =1.0$. Other parameters in all frames: $K=24.0, L=1024$.}
    \label{fig:activewidth}
\end{figure}

To \mcm{quantify} the structure of the filamentous active networks, we compute the mean width of the filaments \mcm{defined as} $w=A_0/\ell_e$, where $\ell_e$ is the total length of the network. 
This 
is \mcm{computed} by ``skeletonizing'' the active component, a process that creates a one dimensional network \mcm{of active region}
by removing, \mcm{over several passes, all} points that do not change its 
\mcm{connectivity} (see \cite{zhang1984fast,SI} for details). We plot the width $w$ as a function of activity $|\alpha|$ and varying surface tension $\sigma$ in Fig.~\ref{fig:activewidth}. In steady state, $w$ does not depend on the composition of the mixture $\phi_0$. Increasing $\phi_0$ only has the effect of creating a denser network, with a larger number of shorter filaments. 
The filament width is thus not set by composition but by activity and surface tension. 
\mcm{The dependence of $w$} on these parameters can be understood through a scaling argument similar to the one we used above for active mixing.
Consider an elongated 
\mcm{active} droplet \mcm{of} width $w$ and length $\ell\gg w$. \mcm{The evolution of the droplet's shape is controlled by the interplay of active flows at rate $\tau_a^{-1} = S(|\alpha|-\alpha_c)/\eta$ that tend to elongate the droplet and diffusive fluxes at rate $\tau_d^{-1} = M\sigma/A^{3/2}$ that tend to make the droplet shape isotropic, where $A \sim \ell w$ is the area of the elongated droplet.
Equating these rates, and using $S\sim \sqrt{(|\alpha|-\alpha_c)/u}$, we estimate  
the width of the network as
$w\sim \sigma^{2/3}/(|\alpha|-\alpha_c)$. This estimate}
is in excellent agreement with numerical measurements over a wide range of values of activity and interfacial tension, as shown in Fig. \ref{fig:activewidth}B.
\mcm{When the passive liquid crystal is in the nematic state
($r<0$), we estimate
$w\sim (\sigma/|\alpha|)^{2/3}$ since $S \sim \pg{\sqrt{-r/u}}$ is independent of activity, which also agrees with} numerical results \cite{SI}.
\\

\textit{\mcm{Discussion.}} We have presented numerical results and an analytical prediction for the phase boundary of the coexistence region of an active liquid crystal and a passive fluid. \mcm{Mixing is driven by activity and 
the location of the boundary is set by the} balance of diffusive and active flows. 
Activity 
completely changes the morphology of the
\mcm{demixed state,}
creating a dynamical network of active material that invades space \mcm{even at the lowest fractions of the active component, with the passive fluid condensing in droplets that fill the interstitial regions.}
\mcm{We stress that the structure of} the dynamical \mcm{active} network 
\mcm{depends} on the dimensionality of space. 
\mcm{Preliminary results (see SI~\cite{SI}) show} that in three dimensions active \mcm{and passive fluids phase separate into a}
bicontinuous structure.

\pg{\mcm{Mixtures of two phase separating polymers with very different relaxation times are also known to form percolating filamentary networks, 
where droplets of the  slow component are trapped in the interstitial regions of the network formed by the fast one~\cite{tanaka1996universality, tanaka2000viscoelastic}. These structures are not, however, stable and at long time always fragment into 
droplets of the minority component.} Activity provides the necessary mechanism to stabilize the network and allows control over the fluid microstructure.}

Our work \mcm{is directly relevant to recent experiments on phase separating mixtures  of active microtubule-based liquid crystals and passive fluids~\cite{tayar2023controlling, zhao2024asymmetric, adkins2022dynamics}. }
It elucidates the mechanisms underlying their complex behavior, which is essential in the path to turn these systems into functional materials.
\\

\begin{acknowledgments}
We thank Zvonimir Dogic, Austin Hopkins, and Liang Zhao for illuminating discussions. This work  was primarly funded by the U. S. Department of Energy Office of Basic Energy Sciences (DE-SC0019733). The work on the characterization of the coexistence line received additional funding from NSF-DMR-2041459 and NSF-DMR-2011846. Use was made of computational facilities purchased with funds from the National Science Foundation (CNS-1725797) and administered by the Center for Scientific Computing (CSC). The CSC is supported by the California NanoSystems Institute and the Materials Research Science and Engineering Center (MRSEC; NSF DMR 2308708) at UC Santa Barbara.

\end{acknowledgments}

\bibliography{ref}

\end{document}